# Electrochemical Cells with Intermediate Capacitor Elements

**Haim Grebel and Akshat Patel**

*The Electronic Imaging Center at NJIT and the Department of Electrical and Computer Engineering,*

*The New Jersey Institute of Technology (NJIT), Newark, NJ 07102. grebel@njit.edu*

**Abstract:** Our goal is to electronically regulate electrochemical cells. For this, we introduced a third element, called the gate, which was placed between the cathode and the anode electrodes of the cell. Voltage applied to this element controlled the local potential of the electrolyte, thus impacting the flow of ions within the cell. The flow of ions was monitored by the electronic current in the external cell's circuit. We provide simulations and experimental data as proof to the validity of this concept. This is but the first step towards a demonstration of a two-dimensional, bi-carrier ion transistors.

Electrochemical cells have been studied since the early eighteenth century [1,2]. Two half-cell reactions may be identified: one is the oxidation of the anode, dissolving positive metal ions into an electrolyte. The generated electrons flow from the anode through an external load to the second half-cell, containing the cathode, where a reduction is taking place; in our case, it is the generation of hydrogen. The circuit is completed by drift and diffusion of ions within the electrolyte(s) in both half-cells. The two-half cells are typically connected by a salt-bridge, which enables passage of ions, yet, limits the flow of the bulk electrolyte molecules. We replaced the salt bridge by an electronic element, which we call the gate. In the past we have used graphene – monolayer of graphite – as a single gate electrode [3,4]. Here we use a capacitor-like gate element, instead.

## 2. Experiment

Simulations employed a commercial tool, which is based on finite elements (COMSOL). A wet-cell, rectangular-shape battery, with Zn electrode as the anode and Pt electrode as the cathode was used. The model took into account the reactions on the anode (oxidation) and on the cathode (formation of hydrogen). The diffusion of ions in the cell considered only excess of $Zn^{2+}$ ions in the electrolyte. The local current was assessed as the negative derivative of the local electrolyte potential multiplied by the electrolyte conductivity. The permeable capacitor was modeled by two porous electrodes without (and sometimes with) a mid- porous insulating film. The effective electrolyte-to-metallic volume ratio in the porous electrode was 1/9. In this ideal case, we did not consider a reaction between the metallic capacitor and the electrolyte, hence, the potential of the solid portion of the porous electrode was kept at the gate potential, Vg. Other simulation parameters were: electrical conductivity of Pt, Zn, porous capacitor electrodes and $Zn^{2+}$ in the electrolyte, respectively, $10^7$, $10^7$, $3x10^5$, 0.01 S/m. The upper tip of the Pt cathode was grounded and the upper tip of the Zn anode was kept at -0.78 V, slightly lower than the standard potential of the Zn anode ($E_0^{(Zn)}=-0.82$ V). This means that the cell's voltage was 0.78 V. While the model is simple, it was able to capture the essence of effect.

In the experiments we used either 5% acetic acid, or 5% $H_2SO_4$ as electrolytes in both half-cells. The $Zn^{2+}$ ions ended up in the electrolyte whereas hydrogen gas was formed near the Pt electrode. The gate electrode was constructed as a permeable capacitor: two thin Au/Pd films were sputtered on either side of a 10-micron Teflon filter. The resistance of each film was ca $2 \times 10^4$ Ohms/cm. The gate capacitor was placed on a plastic plate at the cell's center. A 0.375 $cm^2$ hole in this mid-plate allowed ions to flow through [3,4]. The gate element was, thus, conductive and permeable to the passing ions. One plate of the capacitor facing the anode was grounded; the gate bias Vg was applied to the other plate, facing the cathode (Figure 1a). The Pt cathode was grounded to prevent the biasing potential from floating. A reference electrode was attached to the biased plate with some potential drop at the contact. Calibration curve of the gate potential with respect to the reference electrode was obtained (Figure 1b).

## 3. Results

In Figure 2a,b we show simulations of the electrolyte potential at various cell's cross sections for Vg=±0.1. In Figure 2c,d we show the electrolytic concentration distribution (mol/$m^3$) at the capacitor for Vg=±0.1 V. In Figure 3a we show simulations of the average electrolyte potential at the gate element as a function of the gate bias. The decrease in the electrolyte potential resulted in an electronic current decrease, which was assessed at the Pt electrode upper tip (Figure 3b).

In Figure 4a,b we show experimental data: here, the cell's current, flowing in the outer circuit through a load of 100 Ohms, is plotted as a function of the gate bias (Figure 4a). In Figure

4b we plot the cell's current as a function of the voltage developed between the capacitor and the reference electrode. The curve is shifted due to the standard potential of the Au/Pd film in addition to the small voltage drop at the contact. When the cell's load was changed to 50 KOhms, the I-V curve became non-linear, saturating at V~0 V with respect to the reference electrode. This may be expected if hydrogen is generated at the capacitor plate.

## 4. Discussion and Conclusions

The difference between the potentials on the solid electrode $\phi_S$, the liquid electrolyte $\phi_L$ and the standard potential of the electrode $E_0$ is known as the overpotential: $\eta=\phi_S-\phi_L-E_0$. Initially, at the cathode, $\eta\sim 0$ and the electrolyte potential near the grounded cathode ($\phi_S=0$, $E_0^{(cathode)}=0$) is small; $E_0^{(cathode)}=0$ because hydrogen is formed at the cathode. At the Zn anode, $\phi_S=-0.78$ V and $E_0^{(Zn)}=-0.82$ V, thus the initial electrolyte potential is small, as well. If we assume that the porous gate electrode is screened by the slow flowing ions, then the electrolyte potential at its vicinity will be proportional to $-V_g$ and will decrease upon increasing the gate bias. On the basis of charge distribution, this is how we interpret Figure 3a. Control over the ion flow is translated into control of the electrical current in the external circuit.

Our gate element was of the leaky type allowing ion transport through it. Such internal capacitor appears in parallel to the cell's capacitance, which is formed between the anode and the cathode of the cell; its overall effect is to increase the cell's capacitance. Thus, replacing the metallic capacitor with an electronic barrier, i.e., a *p-n* junction has ramifications to supercapacitors [5].

While the experimental data followed qualitatively the simulations it did not approach the zero current levels as anticipated [3,4]. This is due to several reasons: (1) we did not apply a gate bias beyond ±1 V in order to avoid secondary reactions. Secondary reactions (e.g., hydrolysis or Cl formation when using NaCl as an electrolyte) may limit the overall effect. (2) Parasitic effects near the metal-plated capacitor, such as, electron charge dissipation may also counter the control mechanism. (3) The cell's load needs to be comparable, or smaller than the cell's resistance (measured with opened circuit). This implies that highly conductive electrolytes will require smaller cell's load. Indeed, the cell's characteristics were similar to Figure 4c when using a cell's load resistance of 100 Ohms and 5% $H_2SO_4$ (1 M) as an electrolyte instead of the acetic acid (not shown).

Finally, we note that ion control by use of single charge species have been studied [6]. Here we attempt to control both positive and negative charge types by a single control element. Application are envisioned in electronic control of ion flow and logic ion gates.

**Acknowledgement**

With partial support by NSF/EEC 1343716.

**Figures**

Figure 1. (a) The cell's configuration. (b) Calibration in up and down ramp: capacitor voltage with respect to the reference electrode as a function of applied bias.

Figure 2. Simulations: time dependent solution obtained after 60 minutes: local electrolyte potential (in Volts) when the applied gate voltage was (a) Vg=-0.1 V and (b) Vg=+0.1 V. The cell's voltage was 0.78 V. The electrolyte potential at the rim of the gate element is of no consequence since this portion of the gate electrode is outside the cell. The $Zn^{2+}$ ion concentration distribution (in mol/m3) for (c) Vg=-0.1 V and for (d) Vg=+0.1 V.

Figure 3. Simulations as a function of the gate voltage: (a) the averaged electrolyte potential at the mid-gate element. (b) Cell's current as assessed at the tip of the cathode.

Figure 4. Experiments: (a) cell's current as a function of the gate voltage. The load was 100 Ohms. (b) The same as (a) but the voltage on the capacitor is referenced to Ag/AgCl reference electrode. (c) With cell's load of 50 KOhms. The inset shows the calibration curve in up and down ramp.

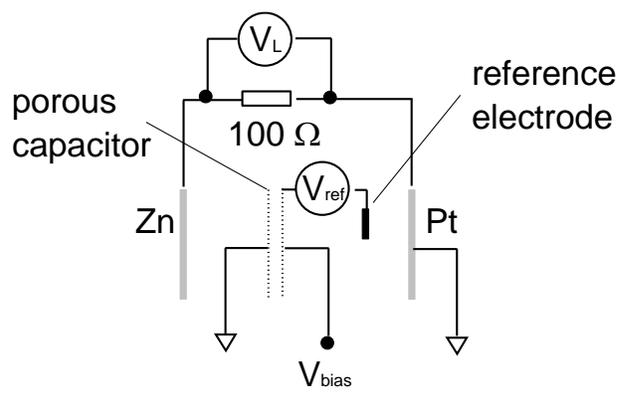 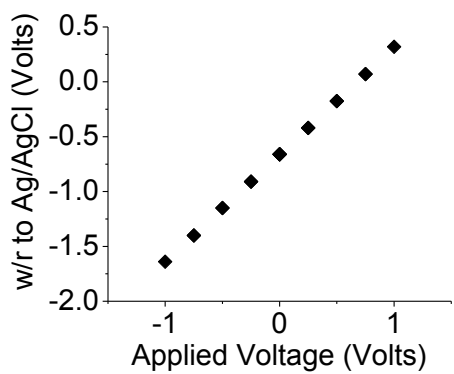

Figure 1. (a) The cell's configuration. (b) Calibration in up and down ramp: capacitor voltage with respect to the reference electrode as a function of applied bias.

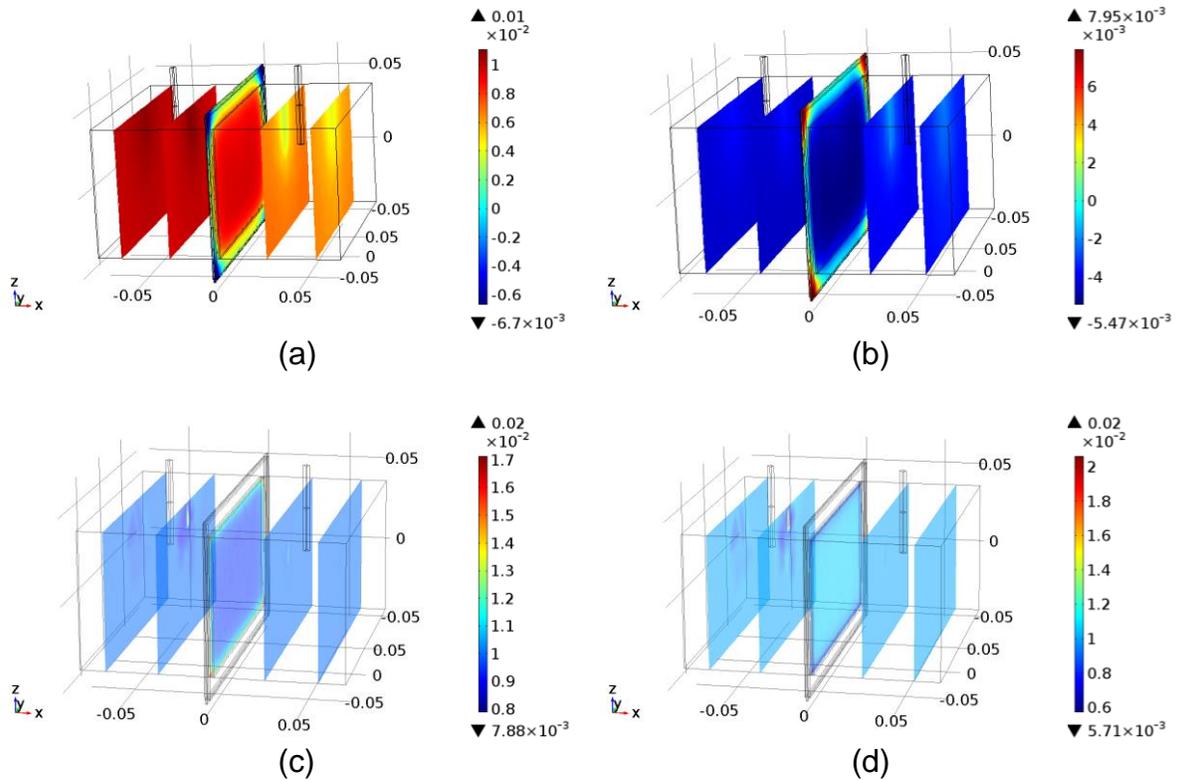

Figure 2. Simulations: time dependent solution obtained after 60 minutes: local electrolyte potential (in Volts) when the applied gate voltage was (a) Vg=-0.1 V and (b) Vg=+0.1 V. The cell's voltage was 0.78 V. The electrolyte potential at the rim of the gate element is of no consequence since this portion of the gate electrode is outside the cell. The $Zn^{2+}$ ion concentration distribution (in mol/m$^3$) for (c) Vg=-0.1 V and for (d) Vg=+0.1 V.

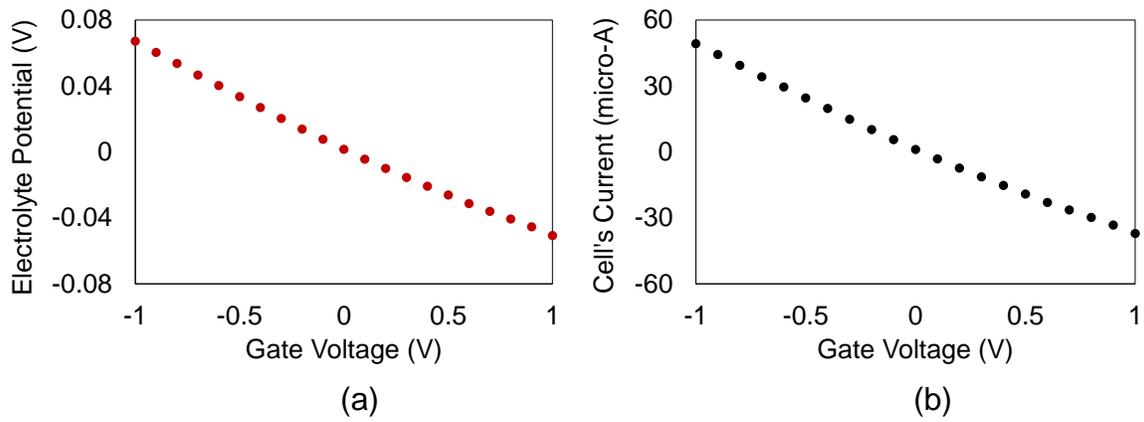

Figure 3. Simulations as a function of the gate voltage: (a) the averaged electrolyte potential at the mid-gate element. (b) Cell's current as assessed at the tip of the cathode.

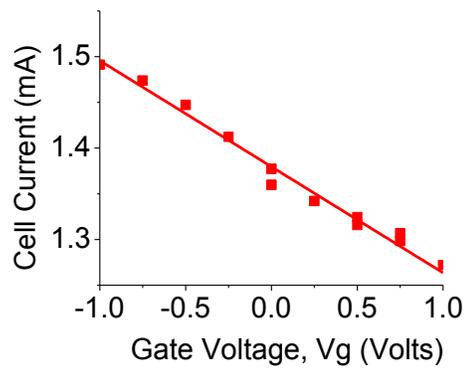 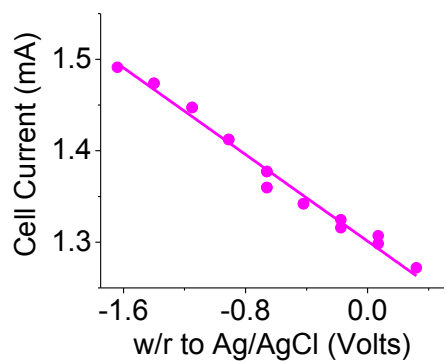

(a)                                 (b)

(c)

Figure 4. Experiments: (a) cell's current as a function of the gate voltage. The load was 100 Ohms. (b) The same as (a) but the voltage on the capacitor is referenced to Ag/AgCl reference electrode. (c) With cell's load of 50 KOhms. The inset shows the calibration curve in up and down ramp.